\RequirePackage{fix-cm}
\documentclass[twocolumn]{svjour3}          
\smartqed  
\usepackage{amsmath}
\usepackage{amsfonts}
\usepackage{amssymb}
\usepackage{dsfont}
\usepackage{graphicx}
\usepackage{natbib}
\usepackage{url}
\begin{document}

\title{Bayesian inference in hierarchical models by combining independent posteriors
\thanks{This work was financially supported by the Academy of Finland (Finnish Centre of Excellence in Computational Inference Research COIN).}
}


\author{Ritabrata Dutta \and
        Paul Blomstedt \and
        Samuel Kaski
}

\authorrunning{Dutta, Blomstedt and Kaski} 

\institute{Ritabrata Dutta \and Paul Blomstedt \and Samuel Kaski \at
              Helsinki Institute for Information Technology HIIT, Department of Computer Science, Aalto University, Espoo, Finland\\
              \email{ritabrata.dutta@aalto.fi, samuel.kaski@aalto.fi}           
}

\date{Received: date / Accepted: date}

\maketitle

\begin{abstract}
Hierarchical models are versatile tools for joint modeling of data sets arising from different, but related, sources. Fully Bayesian inference may, however, become computationally prohibitive if the source-specific data models are complex, or if the number of sources is very large. To facilitate computation, we propose an approach, where inference is first made independently for the parameters of each data set, whereupon the obtained posterior samples are used as observed data in a substitute hierarchical model, based on a scaled likelihood function. Compared to direct inference in a full hierarchical model, the approach has the advantage of being able to speed up convergence by breaking down the initial large inference problem into smaller individual subproblems with better convergence properties. Moreover it enables parallel processing of the possibly complex inferences of the source-specific parameters, which may otherwise create a computational bottleneck if processed jointly as part of a hierarchical model. The approach is illustrated with both simulated and real data.
\end{abstract}

\section{Introduction}
\label{intro}
In problems where data are available from a number of different, but related, sources (e.g. studies, experiments, treatment units, sites of observation), or when the assumed data generating mechanism can be thought of as having a hierarchical structure, it is often natural to formulate the modelling problem in terms of a \emph{hierarchical model}, which in a generic form can be written as
\begin{align}
X_{ji} &\sim F_{X|\theta}(\cdot|\theta_j),\;j=1,\ldots,J, \;i=1,\ldots,n_j \label{eq:obs}\\
\theta_j &\sim F_{\theta|\phi}(\cdot|\phi), \label{eq:param}.
\end{align}
Here the $X_{ji}$ are random variables corresponding to the observed data from source $j$, $\theta_j$ are parameters specific to each of the $J$ data sources, and finally, $\phi$ is a common hyperparameter. In Bayesian inference, a prior distribution 
\begin{equation}
\phi \sim F_{\phi}, \label{eq:hyperparam}
\end{equation}
is additionally specified on $\phi$, which implies an assumption of exchangeability between the parameters $\theta_j$. The overall goal is to make inference about the unknown hyperparameter $\phi$, as well as about the parameters $\theta_j$, given the observed data from all of the $J$ sources. In particular, $\phi$ often has a direct interpretation in terms of a characterization (e.g. central tendency and/or spread) of the distribution from which the $J$ parameters $\theta_j$ are thought to have been generated. Furthermore, integrating out $\phi$ from the joint posterior distribution of $(\theta_1,\ldots,\theta_J,\phi)$ induces dependences between the remaining parameters, such that the marginal posterior distribution for each $\theta_j$ is able to ``borrow strength'' across the hierarchy when the sample sizes $n_j$ are small.  For further details on hierarchical models, see e.g. \cite{Gelman_et_al_2013}.

Since all parameters of a hierarchical model need to be inferred jointly, the computational task may quickly become practically infeasible as the number of sources $J$ grows. Furthermore, the models $F_{X|\theta}(\cdot|\theta_j)$ may be of an arbitrarily complex form, which is only implicit in the generic notation of Equation (\ref{eq:obs}). The models may possibly involve additional parameters, covariates or latent variables, or be hierarchical models themselves, which further exacerbates the computational task. Finally, in addition to the high computational cost, Markov chain Monte Carlo (MCMC) inference schemes for complex and high dimensional hierarchical models have been found to suffer from unfavorable convergence properties, as recently reported by \cite{Rajaratnam_Sparks_2015}. 

In remedy of the above problems, we propose in this paper a parallelization scheme for hierarchical models, where we first perform posterior inference independently for each $\theta_j$, and then, treating the independent posterior distributions as observed data, we use a substitute hierarchical model to replace the original model (\ref{eq:obs})--(\ref{eq:hyperparam}) for joint parameter inference. More specifically, we introduce for each $j$ a parameter $\psi_j$, explicitly defined as the mean parameter of the observed posterior sample, with the implication that $\psi_j$ and $\theta_j$ share the same support. Assuming further that $F_{\psi|\phi}$, the conditional prior distribution on all $\psi_j$, lies in the same parametric family as $F_{\theta|\phi}$ (see Equation (\ref{eq:param})), we use a suitably constructed, scaled likelihood function to make posterior inference about $(\psi_1,\ldots,\psi_J)$, with dependence introduced through $\phi$. The scaling of the likelihood function prevents the posterior marginals on $\psi_j$ from concentrating as the independent posterior sample sizes grow. The resulting joint posterior distribution for $(\psi_1,\ldots,\psi_J,\phi)$ then provides an approximation to the posterior on $(\theta_1,\ldots,\theta_J,\phi)$ in the original hierarchical model. While meta-analysis \citep[e.g.][]{Higgins_2009} is not the main focus of this paper, the proposed approach is essentially based on combining posterior samples from a set of Bayesian analyses in a meta-analysis-like manner. We will therefore refer to the approach later in the paper as \emph{meta-analysis of Bayesian analyses}, see the discussion in Section~\ref{sec:interpretation}.

At first sight it may appear that not much has been gained by replacing the original model by a parallelization step, followed by inference in another hierarchical model. However, the advantages of the proposed approach are twofold. First, it speeds up convergence by breaking down the initial large inference problem into smaller individual subproblems with better convergence properties. In the subsequent hierarchical reconstruction step, convergence is further facilitated by the fact that the independently generated source-specific posterior samples, now treated as data, may themselves already provide a reasonable approximation to the targeted dependent posteriors on $\psi_j$. Second, the proposed approach enables parallel processing of the possibly complex inferences on $\theta_j$, which may otherwise create a computational bottleneck if processed jointly as part of a a hierarchical model. Disregarding for the moment the issue of convergence, assume for concreteness that a Gibbs sampler for a hierarchical model with $J$ data sources requires $K>1$ units of computation to generate a value from the full conditional distribution of $\theta_j$, such that the total number of units in the full model amounts to $KJ+1$ per iteration. In terms of total computational effort, the parallelized model has, under the same assumptions, a cost which is $J$ units higher, $KJ+J+1$, since the inferences on $\theta_j$ are connected using a substitute hierarchical model requiring $J+1$ units per iteration. However, while the entire iteration of $KJ+1$ units in the original model needs to be processed on a single worker, the parallelized model enables distribution of the computations to multiple workers, each processing at most $\max(K,J+1)$ units per iteration. Thus, if computational resources allow for processing $J$ tasks in parallel, the computation time for the parallellized model will be proportional to $K+J+1$, compared to $KJ+1$ for direct inference in the full hierarchical model.

The initial motivation for the work presented in this paper stems from a recent line of research on distributed computational strategies for big data 
\citep{Minsker_etal_2014,Neiswanger_etal_2013,Wang_DDunson_2013,Wang_Peng_Dunson_2014}, where MCMC sampling is first done independently for disjoint subsets of the data, after which the obtained samples are combined to construct a sample from the full-data posterior. The goal of these methods is, however, different from ours and as such, they are not directly applicable to hierarchical models. Another approach which is very similar in spirit to ours has previously been proposed by \cite{Lunn_et_al_2013}. In their work, independently generated posterior samples for the source-specific parameters in a hierarchical model are re-used as proposal values within a Metropolis-Hastings step in learning the full model. This requires that the full samples are stored, and that they are substantial in size to be able compensate for the differences between the independent and full hierarchical posteriors. 
Our strategy is different from the above, in that we specify a model for the posterior mean parameter, where the independent posterior samples enter through a likelihood function, and as a consequence, new posterior samples will be generated from the full hierarchical posteriors. Furthermore, if a suitable parametric form is assumed for the the likelihood function, it is sufficient to store only summaries of the posterior samples.

The rest of the paper is structured as follows. In Section~\ref{methods}, we develop our parallelized approach and discuss some of its theoretical properties. For ease of implementation, we provide analytical forms for the full conditional distributions in two conjugate cases, assuming  $\psi_j$ to be distributed as multivariate normal and inverse Wishart, respectively. In Sections \ref{num_illus} and Section~\ref{sec:cheese}, 
we illustrate our approach in terms of computational efficiency and accuracy of inference with both simulated and real data. 
The paper is concluded with a discussion in Section~\ref{sec:discussion}. 

\section{Methodology}
\label{methods}

Consider the hierarchical model specified in Equations (\ref{eq:obs})--(\ref{eq:hyperparam}). 
When computationally feasible, a standard way of conducting full Bayesian inference in this model is through a Gibbs sampling scheme, in which parameter values are successively sampled as 
\begin{align}
\phi &\sim \pi(\phi|\theta_1,\ldots,\theta_J)\propto \pi(\phi)\prod_{j=1}^J \pi(\theta_j|\phi),\label{eq:gibbs_phi}\\
\theta_j &\sim \pi(\theta_j|\phi,\mathbf{x}_{j})
\propto \pi(\theta_j|\phi)\prod_{i=1}^{n_j} f(x_{ji}|\theta_j)\label{eq:gibbs_theta}\\
&\phantom{\sim  \pi(\theta_j|\phi,\mathbf{x}_{j})~}=\pi(\theta_j|\phi)\,l(\theta_j;\mathbf{x}_{j}),\quad j=1,\ldots,J,\nonumber
\end{align}
where $\pi$ and $f$ denote generic density functions for parameters and data, respectively, $l$ is a likelihood function, and $\mathbf{x}_{j} = (x_{j1},\ldots,x_{jn_j})$ is the observed data from source $j$. By standard arguments, the above sampling scheme produces (correlated) samples under the joint posterior density $\pi(\phi,\theta_1,\ldots,\theta_J|\mathbf{x}_{1},\ldots,\mathbf{x}_{J})$.

As previously motivated in Section~\ref{intro}, we now wish to develop a substitute model to replace the inference in Equations~(\ref{eq:gibbs_phi})--(\ref{eq:gibbs_theta}), using posterior samples of $\theta_j$, generated independently for each $j$ and denoted as $\boldsymbol{\theta}_j^* = (\theta_{j1}^*,\ldots,\theta_{jL}^*)$.
As a first step, we introduce a parameter $\psi_j\triangleq \mathbb{E}(\theta_j|\mathbf{x}_{j})$, with the implication that $\psi_j$ and $\theta_j$ share the same support. Our goal is to replace the hierarchical inference on $\theta_j$ in the original model with that on $\psi_j$. With further justification given in Section~\ref{sec:asymptotics}, we construct a likelihood function
\begin{equation}\label{eq:psi_likelihood}
l^*(\psi_j;\boldsymbol{\theta}_j^*) = \prod_{l=1}^L \left[f^*\left(\theta_{jl}^*|\psi_j\right)\right]^{1/L},
\end{equation}
where $f^*(\cdot|\psi_j)$ is a density function having the same support as the posterior of $\theta_j$, and parametrized by its expectation parameter 
$\psi_j$,
\[
\psi_j \triangleq \int \theta_j f^*\left(\theta_{j}|\psi_j\right) d\theta_j, 
\]
with any other required parameters assumed known or empirically estimated from $\boldsymbol{\theta}^*_j$. Note that the above likelihood function has been scaled to its $L$:th root in order to prevent it from degenerating around its `true' unknown value, as $L\rightarrow\infty$. The choice of $f^*(\cdot|\psi_j)$ will be further discussed in Section~\ref{sec:likelihood_examples}.

Provided that we are able to construct the likelihood as in Equation~(\ref{eq:psi_likelihood}), and that we assume the same conditional prior for $\psi_j$ and $\theta_j$ given $\phi$, i.e.
\begin{equation}\label{eq:prior_assumption}
\pi(\psi_j|\phi) = \pi(\theta_j|\phi)\;\text{when }\psi_j=\theta_j,
\end{equation}
the inference of Equations~(\ref{eq:gibbs_phi})--(\ref{eq:gibbs_theta}) can be approximated with
\begin{align}
\phi &\sim \pi(\phi|\psi_1,\ldots,\psi_J) \propto \pi(\phi)\prod_{j=1}^J \pi(\psi_j|\phi),\label{eq:new_gibbs_phi}\\
\psi_j &\sim \pi(\psi_j|\phi,\boldsymbol{\theta}_j^*)\propto \pi(\psi_j|\phi)\,l^*(\psi_j;\boldsymbol{\theta}_j^*),\;j=1,\ldots,J,\label{eq:new_gibbs_psi}
\end{align}
which generates samples under the joint posterior density $\pi^*(\phi,\psi_1,\ldots,\psi_J|\boldsymbol{\theta}_1^*,\ldots,\boldsymbol{\theta}_J^*)$. Note that steps (\ref{eq:gibbs_phi}) and (\ref{eq:new_gibbs_phi}) in the two sampling schemes are identical apart from the notation of the argument in $\pi(\cdot|\phi)$. 

\subsection{Characterization of the likelihood}\label{sec:asymptotics}

We will next give an approximate characterization of the likelihood function used in the substitute model 
\begin{align*}
&\pi^*(\phi,\psi_1,\ldots,\psi_J|\boldsymbol{\theta}_1^*,\ldots,\boldsymbol{\theta}_J^*)\propto\\
&{\pi(\phi)\prod_{j=1}^J\pi(\psi_j|\phi)\prod_{l=1}^L \left[f^{*}(\theta_{jl}^*|\psi_j)\right]^{1/L}}.
\end{align*}
We begin by expressing the likelihood function $l^*(\psi_j;\boldsymbol{\theta}_j^*)$ as a function of the observed data $(\mathbf{x}_{1},\ldots,\mathbf{x}_{J})$ in the limit, as $L\rightarrow \infty$, and derive the following upper bound for it:
\begin{align*}
l^*(\psi_j;\boldsymbol{\theta}_j^*) &= \prod_{l=1}^L \left[f^{*}(\theta_{jl}^*|\psi_j)\right]^{1/L}\nonumber\\
&=\exp\left[\frac{1}{L}\sum_{l=1}^L{\log \left[f^{*}(\theta_{jl}^*|\psi_j)\right]}\right]\underset{L\rightarrow\infty}{\longrightarrow}\nonumber\\ 
&\phantom{=}~\exp\left[\int \log \left[f^{*}(\theta_{j}|\psi_j)\right]\pi(\theta_j|\mathbf{x}_j)d\theta_j\right]\\
&\leq\log \left[\int f^{*}(\theta_{j}|\psi_j)\pi(\theta_j|\mathbf{x}_j)d\theta_j\right]\\
& = l'(\psi_j|\mathbf{x}_j),
\end{align*}
where the penultimate line is an application of Jensen's inequality and $\pi(\theta_j|\mathbf{x}_j)$ denotes the independent posterior for source $j$. Since $l^*(\psi_j;\boldsymbol{\theta}_j^*)$ is non-negative and 
upper bounded by $l'(\psi_j|\mathbf{x}_j)$, we use the latter as an approximation for the former. Assuming further that $f^{*}(\theta_{j}|\psi_j)$ is sufficiently peaked around the expected value $\psi_j$, we write
\begin{align*}
&l'(\psi_j|\mathbf{x}_j) = \int f^*(\theta_j|\psi_j)\pi(\theta_j|\mathbf{x}_j)d\theta_j\\
&\propto \int f^*(\theta_j|\psi_j)\prod_{i=1}^{n_j}f(x_{ji}|\theta_j)\pi(\theta_j)d\theta_j\\
&\approx \int \delta_{\psi_j}(\theta_j)\prod_{i=1}^{n_j} f(x_{ji}|\theta_j)\pi(\theta_j)d\theta_j\\ 
&= \prod_{i=1}^{n_j}f(x_{ji}|\psi_j)\pi(\psi_j)=l(\psi_j;\mathbf{x}_j)\pi(\psi_j)\\
&=l(\theta_j;\mathbf{x}_j)\pi(\theta_j),
\end{align*}
where $\delta_{\psi_j}$ is the delta function centered on $\psi_j$, and the last equality is due to $\theta_j$ and $\psi_j$ having the same domain, the only change being in the notation of the argument. We therefore have
\[
l^*(\psi_j;\boldsymbol{\theta}_j^*) \approx  l(\theta_j;\mathbf{x}_j)\pi(\theta_j),
\]
where it is seen that the essential difference between difference between $l^*$ and $l$ is in the prior $\pi(\theta_j)$ specified for each source-specific parameter, which conforms with intuition. 

\subsection{Making use of conjugacy}\label{sec:likelihood_examples}

In the sampling scheme of Equations~(\ref{eq:new_gibbs_phi})--(\ref{eq:new_gibbs_psi}) it is often convenient to assume that the posterior distribution for $\phi$ is available in closed form, leaving $\psi_j$ to be sampled using any convenient sampling algorithm. If additionally $l^*(\psi_j;\boldsymbol{\theta}_j^*)$ is assumed to form a conjugate pair with $\pi(\psi_j|\phi)$, the sampling scheme can be even further simplified. While the sampling scheme for the original hierarchical model (see Equations~(\ref{eq:gibbs_phi})--(\ref{eq:gibbs_theta})) may not necessarily be conjugate under this assumption, it enables the utilization of closed-form expressions for efficient sampling in the substitute hierarchical model, provided that $f^*(\cdot|\psi_j)$ gives a reasonably good characterization of the empirical distribution of $\boldsymbol{\theta}_j^*$. Note that conjugacy should be established w.r.t. $\psi_j$, and not all known conjugate pairs retain this property after reparametrization of the likelihood in terms of the expected value. Therefore, care must be taken in the choice of $f^*(\cdot|\psi_j)$. 

An example of inference in a nonconjugate hierarchical model, where conjugacy in the substitute hierarchical model has been utilized, will be given later in Section~\ref{sec:cheese}. In the examples below we use conjugacy to derive the appropriate full conditional distributions for step (\ref{eq:new_gibbs_psi}) in two useful cases: assuming  $\psi_j$ to be distributed as multivariate normal and inverse Wishart, respectively.

\paragraph{\textbf{Example 1} Multivariate normal.} 

The most straightforward application of conjugacy emerges when the parameters $\theta_j$ are \emph{a priori} assumed to have $p$-dimensional multivariate normal distributions. To denote this, we write  
$\phi = (\mu,\mathrm{\Sigma})$ and $\theta_j\sim \mathcal{N}_p\left(\mu,\mathrm{\Sigma}\right)$. 
We then require that $\psi_j\sim \mathcal{N}_p\left(\mu,\mathrm{\Sigma}\right)$, as implied by assumption (\ref{eq:prior_assumption}). 
Conjugacy is further attained by assuming that $\theta_{jl}^*\sim \mathcal{N}_p(\psi_j,\mathrm{S}_j)$, $l=1,\ldots,L$, where the covariance matrix $\mathrm{S}_j$ is assumed to be known but may in practice be estimated from $\boldsymbol{\theta}_j^*$. From the above assumptions and by application of Equation~(\ref{eq:psi_likelihood}), the full conditional distribution in Equation (\ref{eq:new_gibbs_psi}) can then be written as
\begin{align}\label{eq:MBA_Normal_posterior}
\psi_j &\sim \mathcal{N}_p\left(\mu',\mathrm{\Sigma}'\right),\quad j = 1,\ldots,J,\\
&\phantom{\sim~} \mathrm{\Sigma}' = (\mathrm{\Sigma}^{-1}+\mathrm{S}_j^{-1})^{-1}\nonumber\\
&\phantom{\sim~} \mu' = \mathrm{\Sigma}'\left(\mathrm{\Sigma}^{-1}\mu +\mathrm{S}_j^{-1}\frac{1}{L}\sum_{l=1}^{L}\theta_{jl}^{*}\right).\nonumber
\end{align}

\paragraph{\textbf{Example 2} Inverse Wishart.} 
Suppose that we wish to apply the sampling scheme (\ref{eq:new_gibbs_phi})--(\ref{eq:new_gibbs_psi}) for inference in the model $\mathrm{\Theta}_j\sim \mathcal{W}_p^{-1}(\mathrm{\Phi},\kappa)$, i.e. a $p$-dimensional inverse Wishart distribution with scale matrix $\mathrm{\Phi}$ and degrees of freedom $\kappa$. This is a natural choice if the parameter of interest $\mathrm{\Theta}_j$ is a positive definite matrix. In the canonical case of the observed data $\boldsymbol{\mathrm{X}}_j$ having a multivariate normal distribution with an unknown covariance matrix $\mathrm{\Theta}_j$ and fixed mean, the posterior distribution remains within the inverse Wishart family by conjugacy. Another instance of conjugacy results if the data are Wishart distributed with scale matrix $\mathrm{\Theta}_j$ and fixed degrees of freedom.  

As above, we require that $\mathrm{\Psi}_j\sim \mathcal{W}_p^{-1}(\mathrm{\Phi},\kappa)$. 
If we now wish to exploit conjugacy, we must choose a suitable family for $f^*(\cdot|\mathrm{\Psi}_j)$, having the same support as $\pi(\mathrm{\Theta}_j|\boldsymbol{\mathrm{X}}_j)$, and remaining conjugate for $\mathcal{W}_p^{-1}(\mathrm{\Phi},\kappa)$ after reparametrization in terms of $\mathrm{\Psi}_j$. These criteria are satisfied if we set $\mathrm{\Theta}_{jl}^*\sim \mathcal{W}_p(\mathrm{\Psi}_j/\nu_j,\nu_j)$, $l=1,\ldots,L$, a Wishart distribution with scale matrix $\mathrm{\Psi}_j/\nu_j$ and degrees of freedom $\nu_j$. When the data sample sizes $n_j$ are available, we may simply set $\nu_j=n_j$. Finally, applying Equation~(\ref{eq:psi_likelihood}) and combining with the prior, we find that the full conditional distribution to be used in sampling step (\ref{eq:new_gibbs_psi}) can be written as 
\begin{align}\label{eq:MBA_Wishart_posterior}
\mathrm{\Psi}_j &\sim \mathcal{W}_p^{-1}\left(\mathrm{\Phi}',\kappa'\right),\quad j = 1,\ldots,J,\\
&\phantom{\sim~} \mathrm{\Phi}' = \mathrm{\Phi} + \nu_j\frac{1}{L}\sum_{l=1}^{L}\mathrm{\Theta}_{jl}^{*}\nonumber\\
&\phantom{\sim~} \kappa' = \kappa+\nu_j\nonumber.
\end{align}

\subsection{Connection to meta-analysis}\label{sec:interpretation}
The idea of learning a hierarchical Bayesian model on the estimated parameters of related models is commonly used in random effects meta-analysis 
\citep{Burr_Doss_2005,Higgins_2009}, where the data sets themselves are not used and only parameter estimates $\hat{\theta}_j=\hat{\theta}_j(\mathbf{x}_j)$, $j=1,\ldots,J$, estimated from the data of each source $j$ are known. Assuming a single model-family $F_{\theta|\psi}$ for each source $j$, 
\begin{align*}\label{eq:MA}
\hat{\theta}_j &\stackrel{iid}{\sim} F_{\theta|\psi}(\cdot|\psi_j),\;j=1,\ldots,J\\ 
\psi_j &\sim F_{\psi|\phi}(\cdot|\phi)\\
\phi   &\sim F_{\phi},
\end{align*}
the parameters $\psi_j$ in the meta-analysis model are referred to as \emph{random effects}. Generally, $\hat{\theta}_j$ are assumed normally distributed by asymptotic arguments, such that the random effect $\psi_j$ is the mean parameter of the sampling distribution of $\hat{\theta}_j$. Our approach is very similar in spirit, but here we assume $\psi_j$ to be the mean parameter of the source-specific posterior samples, and do not constrain $F_{\theta|\psi}$ to be a normal distribution. Nonetheless, to highlight the similarity, and to convey the idea of conducting meta-analysis on a set of independent Bayesian analyses, we refer to our approach as \emph{meta-analysis of Bayesian analyses} (MBA).

\section{Simulated examples}
\label{num_illus}
In this section, we first compare the MCMC convergence (see details below) for MBA (Equations~(\ref{eq:new_gibbs_phi})--(\ref{eq:new_gibbs_psi})) with direct inference in the full hierarchical model (FHM; Equations~(\ref{eq:gibbs_phi})--(\ref{eq:gibbs_theta})) for Examples 1 and 2 (Section~\ref{sec:likelihood_examples}) as well as a variance components model, which is essentially a combination of the two examples. 
Throughout the examples of this section, conjugacy will be utilized for both MBA and FHM. 

\subsection{Scalability of Inference}
For a fair comparison of the two approaches, we run MCMC iterations until the distribution converges to the stationary distribution, assessed using the Raftery-Lewis diagnostic criterion \citep{Raftery_Lewis_1992}. Here we use this criterion for the posterior mean with confidence bound .05 and probability .95 over 10 pilot runs for both FHM and each step of MBA. This gives us an estimate of burn-in and number of Gibbs iterations required. We report the average computation times up to the point when the Raftery-Lewis criterion is fulfilled. For MBA, the source-specific posteriors are initially sampled independently, each on a different processor. Hence, the time superiority gained by MBA in the following examples is due both to the parallelizability of the MBA algorithm and faster convergence of each of the source-specific parameters.

\paragraph{\textbf{Example 1} (cont.)} 
We consider here a univariate case, with the hierarchical model specified as 
\begin{align*}
X_{ji}&\sim \mathcal{N}(\theta_j,1),\;j=1,\ldots,J, \;i=1,\ldots,n_j\\
\theta_j&\sim \mathcal{N}(\mu,\sigma^2).
\end{align*}
Priors on the hyperparameters are further specified as 
\[
\mu \sim \mathcal{N}(\mu_0,\sigma^2_0),\quad\sigma^2 \sim \mathcal{W}^{-1}_1(\mathrm{\Omega},k).
\]

For inference in FHM, we set $\mu_0 = 0, \sigma^2_0 = 1, \mathrm{\Omega} = 1, k = 3$ \emph{a priori}, and iteratively apply sampling steps (\ref{eq:gibbs_phi})--(\ref{eq:gibbs_theta}) to draw samples from the joint posterior distribution of $(\mu,\sigma^2,\theta_1,\ldots,\theta_J)$, making use of standard conjugate forms. To implement MBA for this model, we first draw posterior samples $\boldsymbol{\theta_j}^*$ independently for each source $j$ using fixed values $\mu=0$ and $\sigma^2 = 1$. Then, having these samples available, we simply substitute the sampling step (\ref{eq:gibbs_theta}) in the FHM sampling scheme with step (\ref{eq:new_gibbs_psi}) in the MBA sampling scheme (steps (\ref{eq:gibbs_phi}) and (\ref{eq:new_gibbs_phi}) are identical and remain unchanged), using the conjugate form given in Equation~(\ref{eq:MBA_Normal_posterior}).

We ran both FHM and MBA for 100 randomly generated data sets with $J = 10$ sources with $n_j=5$ samples in each of them. The true generating values of the parameters for the hierarchical model were $\mu = 2$ and $\sigma^2 = 3$. To compare the posterior distributions inferred by the two approaches, we report in Table~\ref{Comp_example1} for $\phi = (\mu,\sigma^2)$ the coverage probability of the 95\% credible interval (CI) of the posterior distribution, i.e. the proportion of simulations for which the CI covers the true parameter value, the mean squared error (MSE) of the posterior mean estimate, and finally, the average computation time.
\begin{table}
\caption{Comparison of MSE of the posterior mean estimate, coverage probability of 95\% C.I. of the posterior distribution, and average computation time for inference on $\phi = (\mu,\sigma^2)$ using FHM and MBA, respectively, for Example 1 with true values $\mu = 2, \sigma^2 = 3$, and $n_j = 5, J = 10$.}
\label{Comp_example1}       
\begin{tabular}{p{.1\textwidth}p{.13\textwidth} p{.1\textwidth} p{.07\textwidth}}
\hline\noalign{\smallskip}
Method & Avg. computation time & Coverage probability for $(\mu,\sigma^2)$ & MSE \\ 
\noalign{\smallskip}\hline\noalign{\smallskip}
FHM & 0.49 s. & (0.87, 0.87) & 1.68 \\ 
MBA & 0.29 s. & (0.88, 0.91) & 1.69 \\ 
\noalign{\smallskip}\hline
\end{tabular}
\end{table}
We notice that MBA has a similar MSE and coverage probability for each of the parameters and it is almost twice as fast on average than the standard FHM sampling scheme. 

\begin{figure}[h]
    \centering  
            \includegraphics[width=.45\textwidth]{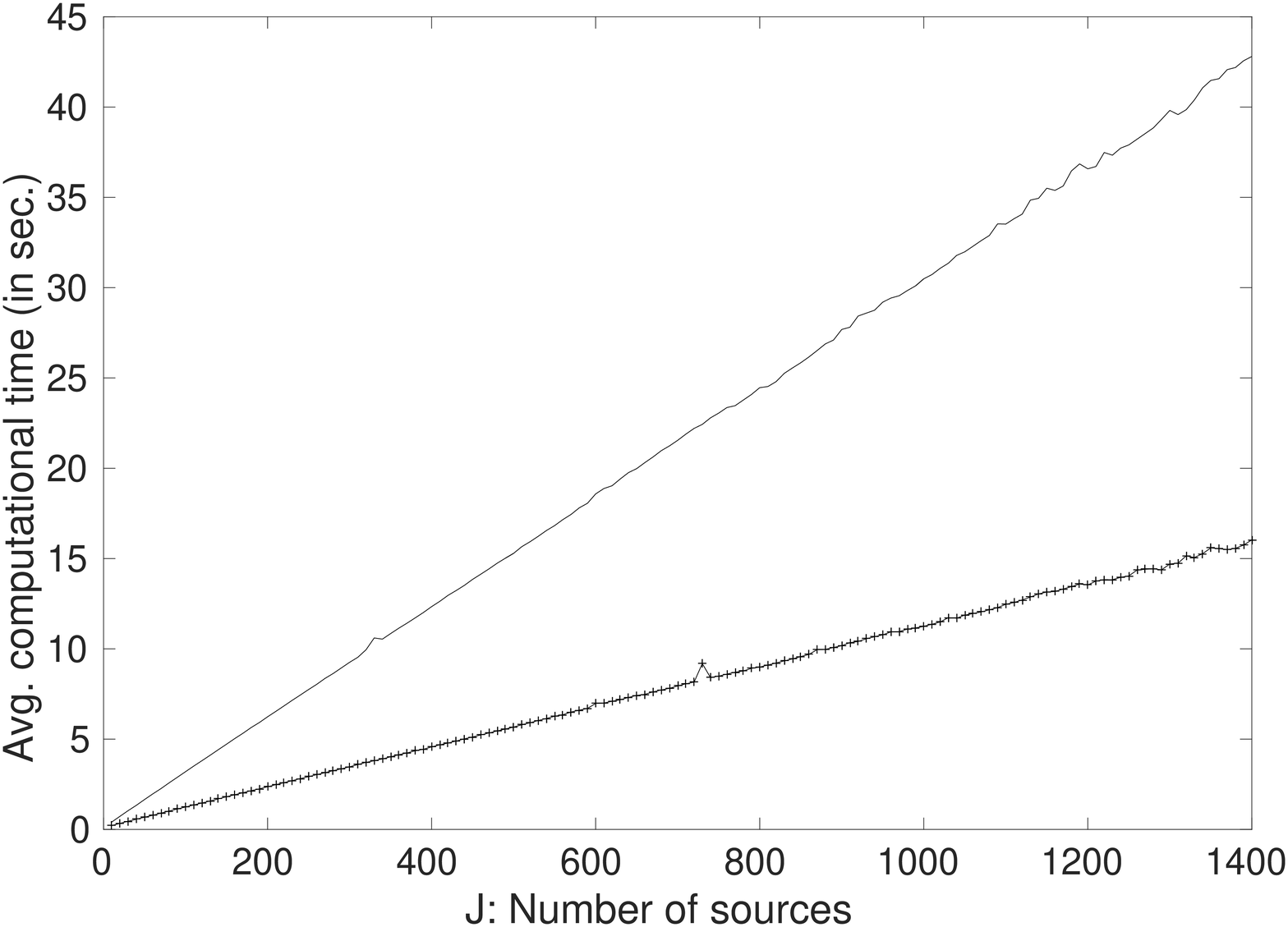}
    \caption{Average computation times for FHM (solid) and MBA (dashed) inference schemes, for Example 1 with $n_j= 5$. }
\label{example1_J}
\end{figure}

Further, to demonstrate the computational advantage of MBA, we report in Figure~\ref{example1_J} the average computation times of FHM and MBA for an increasing number of sources $J$, with fixed $n_j = 5$. As expected, the computational gain is seen to grow linearly with the number of sources.

\paragraph{\textbf{Example 2} (cont.)} 

In this example, we consider inference in the hierarchical model
\begin{align*}
X_{ji}&\sim \mathcal{N}_p(0,\mathrm{\Theta}_j),\;j=1,\ldots,J, \;i=1,\ldots,n_j\\
\mathrm{\Theta}_j &\sim \mathcal{W}_p^{-1}(\mathrm{\Phi},\kappa),
\end{align*}
with a prior on $\mathrm{\Phi}$ specified as
\[
\mathrm{\Phi} \sim \mathcal{W}_p(\mathrm{V},m).
\]
Here, results are demonstrated for $p=1$, $V=1$ and $m = 3$. Standard posterior inference for this model (FHM) proceeds by using the sampling scheme (\ref{eq:gibbs_phi})--(\ref{eq:gibbs_theta}). As before, implementing MBA for the model assumes that we have available, for each source $j$, independently generated posterior samples of, say, size $L$. These were generated using fixed hyperparameter values $\mathrm{\Phi} = 1, \kappa = 3$.  
We then used Equation~(\ref{eq:MBA_Wishart_posterior}) to draw samples in step (\ref{eq:new_gibbs_psi}) of the MBA sampling scheme.

Posterior distributions of $\mathrm{\Phi}$ inferred using the FHM and MBA sampling schemes, respectively, are compared for $n_j = 5,10,20$, with $J = 20$. Values averaged over 100 randomly generated data sets, with true values $\mathrm{\Phi} = 40$ and $\kappa = 3$, are reported in Table~\ref{table:Comp_example2}. Note that the average computation times are not significanlty affected by changes in $n_j$. 
\begin{table}[h]
\caption{Comparison of MSE of the posterior mean estimate, coverage probability of 95\% C.I. of the posterior distribution, and average computation time for inference on $\mathrm{\Phi}$ using FHM and MBA, respectively, for Example 2 with true value $\mathrm{\Phi} = 40$, and $n_j = 5,10,20$, $J = 20$.}
\label{table:Comp_example2}
\begin{tabular}{p{.075\textwidth}p{.075\textwidth}p{.06\textwidth} p{.09\textwidth} p{.075\textwidth}}
\hline\noalign{\smallskip}
Method & Avg. computation time & $n_j$   & Coverage probability & MSE \\ 
\noalign{\smallskip}\hline\noalign{\smallskip}
FHM & 0.42 s.  &$5$   & 0.97 & 40.43  \\
 &~"&$10$ &  0.97 & 29.76   \\
  &̃~"& $20$ &  0.99 & 26.23  \\ 
  &~"& $30$ &  0.98 & 25.21  \\ 
  \hline\noalign{\smallskip}
MBA & 0.17 s.  &$5$ &   0.92 & 54.87 \\
   &~"&$10$ & 0.98 & 35.11   \\
   &~"&$20$ & 1.00 & 28.19   \\ 
   &~"&$30$ & 1.00 & 25.63  \\ 
   \hline\noalign{\smallskip}
\end{tabular}
\end{table}

\begin{figure}[h]
    \centering  
            \includegraphics[width = .45\textwidth]{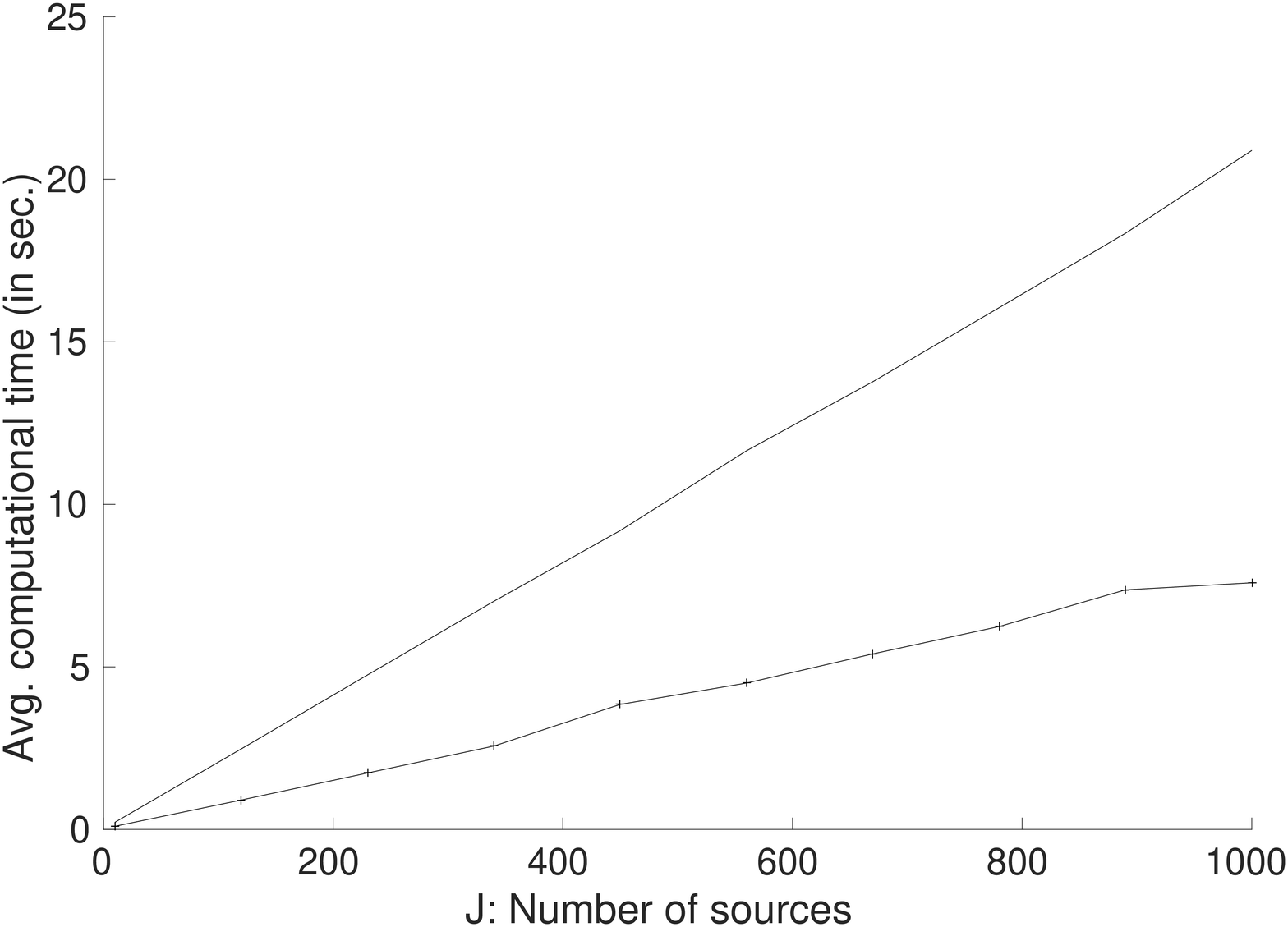}
      \caption{Average computation times for FHM (solid) and MBA (dashed) inference schemes, for Example 2 with $n_j= 10$.}
\label{fig:example2_J_10}
\end{figure}

A comparison of average computation times between FHM and MBA is displayed in Figure \ref{fig:example2_J_10} for an increasing number of sources $J$ and $n_j = 10$.

\paragraph{\textbf{Example 3} Variance components model.}
Next we consider a simple variance components model \citep[see e.g.][]{Goldstein_2011}, specified as
\begin{align*}
y_{ji} = &\beta_0+u_j+v_{ji},\ j=1,\ldots,J,\ i=1,\ldots,n_j\\
&u_j \stackrel{iid}{\sim} \mathcal{N}(0,\sigma_u^2),\ v_{ji} \stackrel{iid}{\sim} \mathcal{N}(0,\sigma_v^2),  
\end{align*} 
where we want to infer the parameters $(\beta_0,\sigma_u^2,\sigma_v^2)$. A reformulation of the above model to better suit our purposes, is to express it in terms of a hierarchical model,
\begin{equation}\label{eq:VC_hier}
Y_{ji} \sim \mathcal{N}(\theta_j,\sigma^2_v),\;\theta_j \sim \mathcal{N}(\beta_0,\sigma^2_u).
\end{equation}
For inference on $(\beta_0,\sigma^2_u)$ using MBA, we first draw samples of $\theta_j$ individually for each $j$, with fixed values $\beta_0 = 0, \sigma^2_u = 1$. We then proceed as in Example 1, with priors on the parameters of interest specified as $\beta_0 \sim \mathcal{N}(0,1),\ \sigma_u^2 \sim \mathcal{W}^{-1}_1(1,3)$. 

The third parameter of interest, $\sigma_v^2$, is not a hierarchical parameter, and could therefore in principle be directly inferred by using all of the data together. However, we will here impose a hierarchical structure also on this parameter, and make inference using MBA. This exemplifies a scenario, where an analyst performing a meta-analysis of $J$ studies is only given posterior samples for the parameters of interest, but not the original data. Thus, we make a slight modification to (\ref{eq:VC_hier}) and specify the model as 
\begin{align*}
Y_{ji} &\sim \mathcal{N}(\theta_j,\tau^2_j)\\ 
\theta_j &\sim \mathcal{N}(\beta_0,\sigma^2_u),\;\tau_j^2 \sim \mathcal{W}_1^{-1}\left(\sigma_v^2(\kappa-2),\kappa\right),
\end{align*}
where the $\tau^2_j$ are now the source-specific variances and $\sigma_v^2$ is the population mean of $(\tau^2_1,\ldots,\tau^2_J)$.   
We then proceed as in Example 2, initially generating posterior samples $\boldsymbol{\tau}^{2*}_j$ independently for each $j$, with hyperparameter values fixed at $\sigma_v^2 = 1, \kappa = 3$. For hierarchical inference using Equation~(\ref{eq:MBA_Wishart_posterior}) we set $\mathrm{\Phi} = \sigma_v^2(\kappa-2)$, and further assume the prior $\mathrm{\Phi} \sim \mathcal{W}^{-1}_1(1,3)$ for posterior inference on $\mathrm{\Phi}$, from which the posterior distribution for $\sigma^2_u = \mathrm{\Phi}/(\kappa-2)$ is then directly obtained. 

\begin{figure}[h!]
    \centering  
            \includegraphics[width = .45\textwidth]{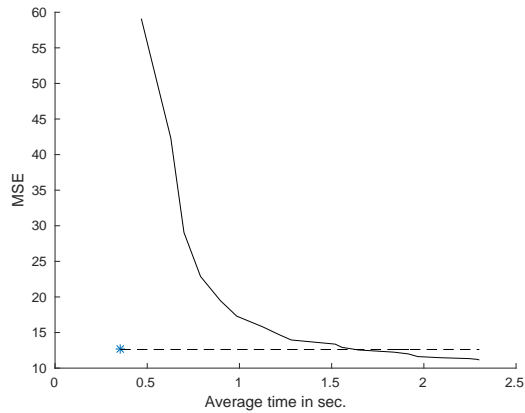}
    \caption{Comparison of computational complexity of FHM (solid) and MBA (star) for Example 3, when $\beta = 30, \sigma_u^2=10, \sigma_e^2=40, J= 15  \mbox{ and } n_j = 18$. The dashed line is included to facilitate comparison with FHM.}
\label{fig:VC_18}
\end{figure}

\begin{figure}[h!]
              \centering
             \includegraphics[width = .45\textwidth]{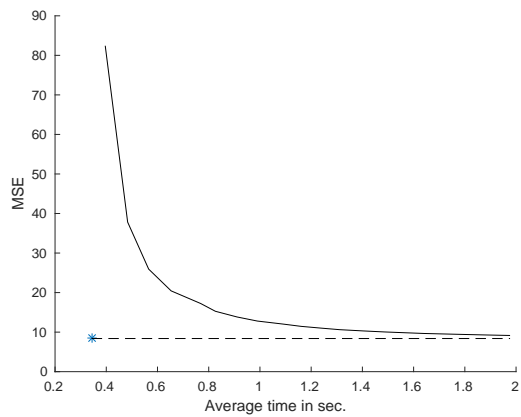}
    \caption{Comparison of computational complexity of FHM (solid) and MBA (star) for Example 3, when when $\beta = 30, \sigma_u^2=10, \sigma_e^2=40, J= 15  \mbox{ and } n_j = 30$. The dashed line is included to facilitate comparison with FHM.}
\label{fig:VC_30}
\end{figure}

For FHM, inference is done using a Gibbs sampling scheme, as described by \cite{Seltzer_et_al_1996}. We report MSE, coverage probability of the 95\% CI, and average computation times in Table~\ref{table:Comp_VC}, when the true generating parameter values are $\beta_0 = 30, \sigma_u^2=10, \sigma_v^2=40$, with $n_j=18$ for different numbers of sources $J=6,12,15$. 
\begin{table}[h]
\caption{Comparison of FHM and MBA on the variance components model in terms of average computation time, coverage probability and MSE, with true parameter values $\beta_0 = 30, \sigma_u^2=10, \sigma_v^2=40$, and $n_j=18$, $J=6,12,15$.}
\label{table:Comp_VC}
\begin{tabular}{p{.03\textwidth}p{.05\textwidth}p{.09\textwidth}p{.13\textwidth}p{.06\textwidth}}
\hline\noalign{\smallskip}
$J$ & Method &  Avg. computation time & Coverage probability ($b_0,\sigma_u^2,\sigma_v^2$) & MSE   \\ 
\noalign{\smallskip}\hline\noalign{\smallskip}
6 & FHM & 1.13 s. & (0.89, 0.85, 0.87) & 18.10 \\
& MBA & 0.23 s. & (0.93, 1.00, 0.92) & 20.46 \\ 
\noalign{\smallskip}\hline
12 & FHM & 1.82 s. & (0.90, 0.92, 0.89) & 10.90 \\
& MBA & 0.30 s. & (0.96, 1.00, 0.91) & 12.14 \\ 
\noalign{\smallskip}\hline
15 & FHM & 1.96 s. & (0.90, 0.85, 0.88) & 11.14 \\
& MBA & 0.33 s. & (0.94, 0.85, 1.00) & 12.26 \\ 
\noalign{\smallskip}\hline
\end{tabular}
\end{table}
To further illustrate the significant improvement in time complexity of MBA over the direct inference, we compute the MSE of the posterior mean estimate from MCMC samples as they get sampled over time. In Figures \ref{fig:VC_18} and \ref{fig:VC_30}, the star indicates (avg. comutation time, MSE) of MBA and the solid curve indicates the improvement of MSE over time for FHM. Note that since the number of burn-in iterations and independently generated posterior samples was fixed in advance, the result for MBA is only shown as one point, whereas for FHM the MSE is computed for an accumulated number of posterior samples after burn-in. An additional dashed line is included to facilitate comparison between the MSE levels of the two approaches.  

\section{Retail sales data}
\label{sec:cheese}
Here we consider a large data set of weekly sales volumes of sliced cheese in 88 retail stores, first used by \cite{Boatwright_1999}. Following \cite{Braun_Damien_2015}, we assume that the sales volume $V_{jt}$ of the $j$:th store in the $t$:th week has a gamma distribution with mean $\lambda_{jt}$ and shape parameter $r_j$. Further, we treat the mean $\lambda_{jt}$ as a log-linear function of the logarithm of the cheese price $P_{jt}$ and the percentage of volumes on display $D_{jt}$ that week, 
\[
\log(\lambda_{ji}) = \beta_{j1}+\beta_{j2}\log (P_{ji})+\beta_{j3} D_{ji}.
\]
A half-Cauchy prior \citep{Gelman_2006} with scale parameter 5 on $r_j$, and a multivariate normal prior on $\boldsymbol{\beta}_j$ with mean $\boldsymbol{\mu}$ and covariance matrix $\mathrm{\Sigma}$ gives us the following hierarchical model 
\begin{align*}
%
V_{jt} &\sim \mathrm{Gamma}(r_j,r_j/\lambda_{jt}),\quad j=1,\ldots,J,\;t = 1,\ldots,T_j,\\
\boldsymbol{\beta}_j &\sim \mathcal{N}_3(\boldsymbol{\mu}, \mathrm{\Sigma}),\quad r_j \sim \text{Half-Cauchy}(5).
\end{align*}
Further, a weakly informative normal-inverse-Wishart prior on $(\boldsymbol{\mu},\mathrm{\Sigma})$ is assumed. 

This example is significant for two reasons, namely, the large number of sources $(J=88)$ which creates a bottleneck for standard MCMC sampling schemes, and the absence of conjugacy in the above hierarchical model. For Bayesian inference in FHM, we use Hamiltonian Monte Carlo \citep[HMC;][]{Neal_2011} implemented in the Stan software package (Stan Development Team 2014). We ran 4 parallel chains for 1000 iterations after a burn-in of the same number of iterations. As the samples show very little correlation, we consider this to be a good estimate of the posterior distribution. The 1000 iterations of Stan for the above full hierarchical model takes around 1 hour to run. 

To implement MBA, we first draw 1000 posterior samples of $\boldsymbol{\beta}_j$ for each source $j$ using Stan, which takes 5 s. on average per source. Using these source-specific posterior samples $\boldsymbol{\beta}^*_j$, 
we apply Equation~(\ref{eq:MBA_Normal_posterior}) for conjugate hierarchical inference on the source-sepcific parameters, and standard conjugate forms for inference on $(\boldsymbol{\mu},\mathrm{\Sigma})$, with priors specified as 
\[
\boldsymbol{\mu} \sim \mathcal{N}_3(\boldsymbol{0},I_3),\ \mathrm{\Sigma} \sim \mathcal{W}^{-1}_3(I_3,6),
\]
where $I_3$ denotes the $3\times 3$ identity matrix. For MBA, a Gibbs sampler to draw 1000 posterior samples from the parameters of interest, after throwing away 1000 burn-in iterations, takes 5 s. Hence, if computational resources are available for parallel computing, using MBA we can gain a 300-fold computational advantage over FHM. 

\begin{figure}[h]
              \centering
             \includegraphics[width = .45\textwidth]{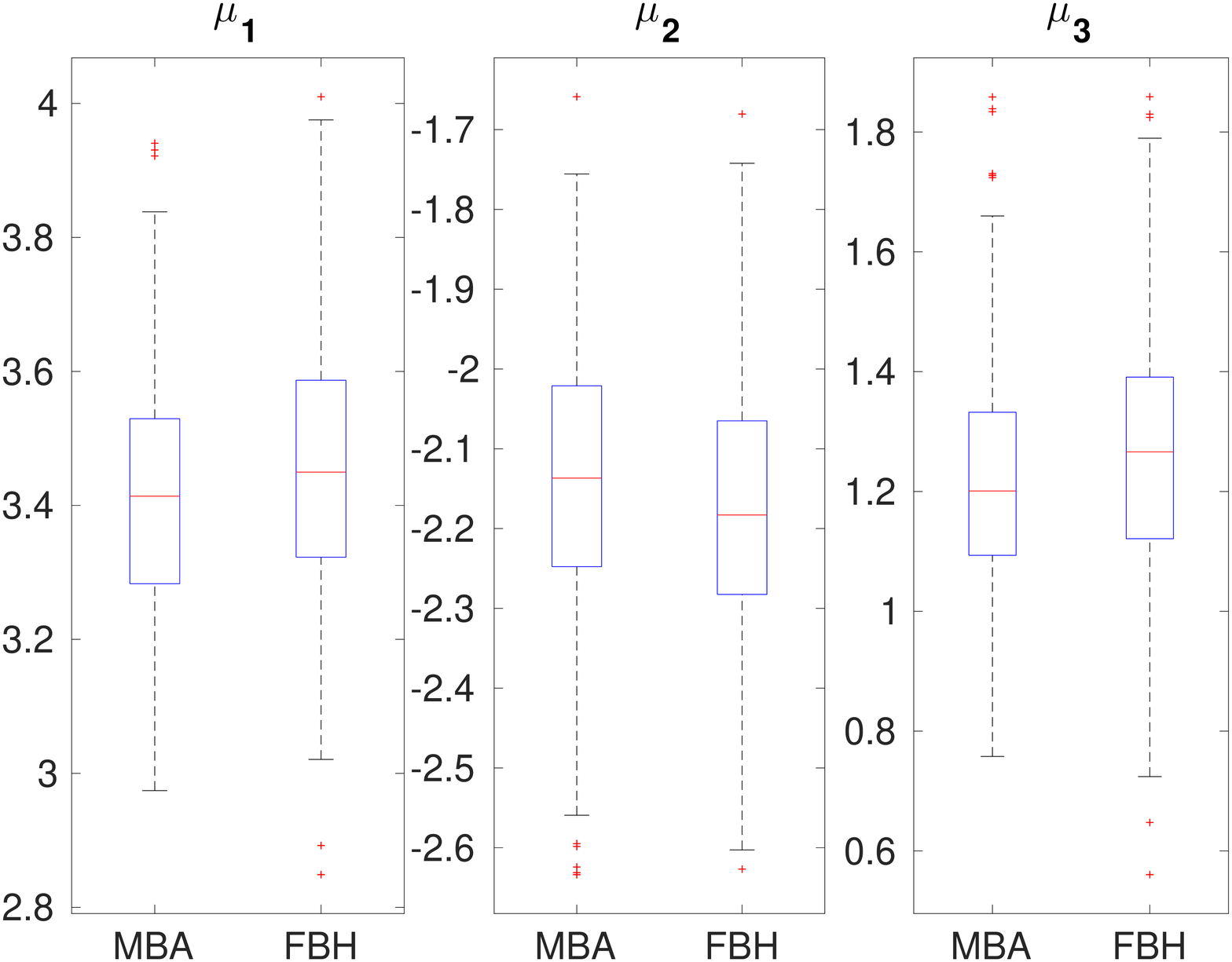}
      \caption{Posterior distributions of $\mu_1,\mu_2,\mu_3$ from MBA and FHM, on retail sales data.}
\label{fig:cheese_mu}
\end{figure}

\begin{figure}[h]
              \centering
             \includegraphics[width = .45\textwidth]{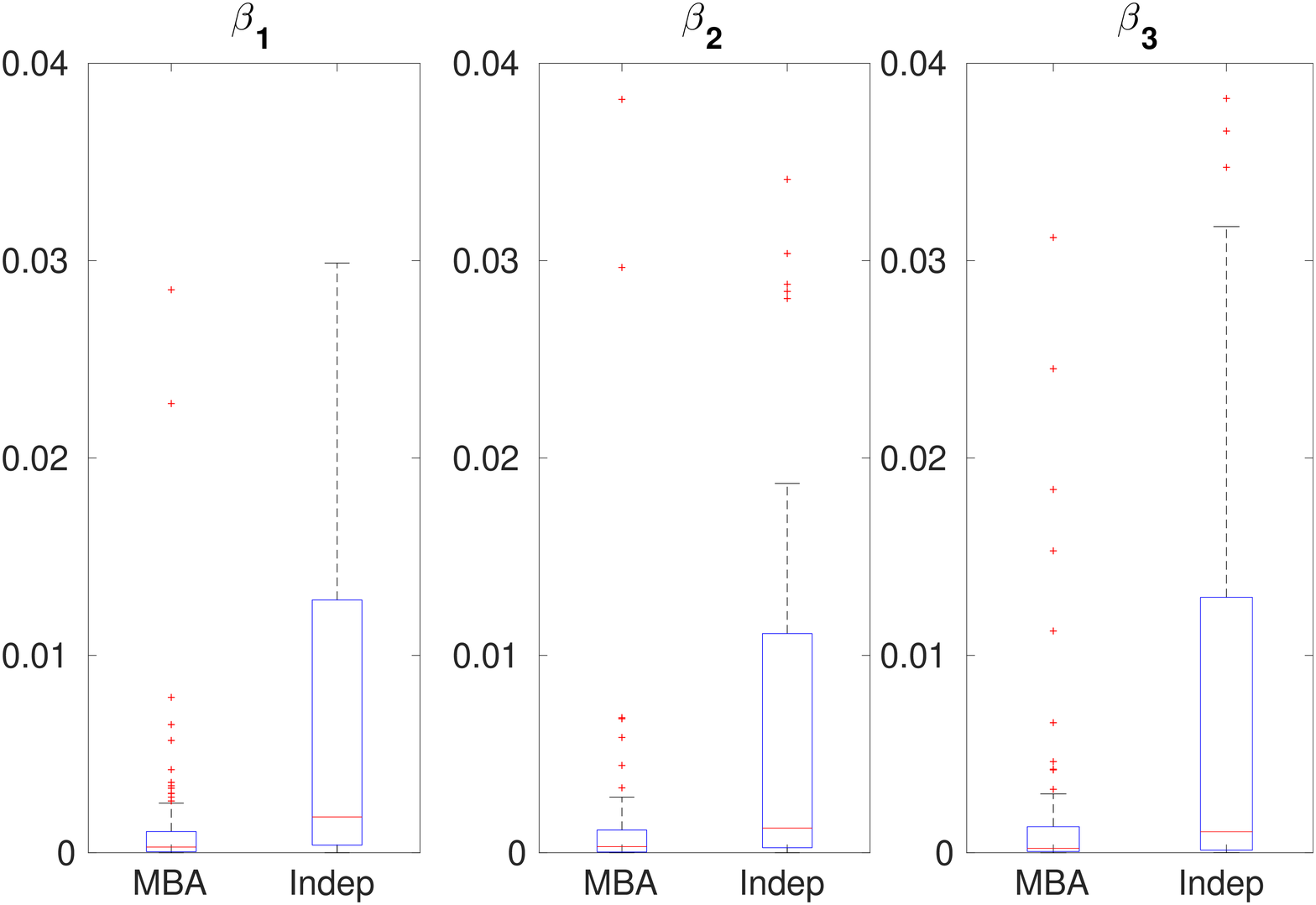}
      \caption{Squared differences of MBA and independent posterior mean estimates, repsectively, from FHM posterior mean estimates of ($\beta_{j1},\beta_{j2},\beta_{j2}$), on retail sales data.}
\label{fig:cheese_beta}
\end{figure}

In Figure~\ref{fig:cheese_mu}, we first compare the posterior distributions of the parameters $\mu_1,\mu_2,\mu_3$ from FHM using HMC with those obtained using MBA. Further, we computed for each source $j$, the squared difference between the posterior means of ($\beta_{j1},\beta_{j2},\beta_{j2}$) estimated from FHM, and those estimated from MBA and the independent posteriors, repsectively. The distributions of these differences, over the 88 sources, are shown in Figure
~\ref{fig:cheese_beta}. Due to the dependence introduced by MBA, the posterior mean estimates are on average much closer to the full hierarchical posteriors than those estimated from the independent posteriors.

\section{Discussion}\label{sec:discussion}

We have presented an approach for parallelized Bayesian inference in hierarchical models, where inference is first conducted independently for the parameters of each data set, after which the obtained posterior samples are used as observed data in a substitute hierarchical model, which is an approximation to the original model. We have shown computational advantages of this approach over direct inference in full hierarchical models, while achieving good accuracy in terms of truthfulness to the original model. Since the total computational effort is slightly higher for the proposed approach than for direct inference (see Section~\ref{intro}), the biggest computational gains can be expected for models with complex source-specific models or a large number of sources, where posterior computations for the source-specific components can be done in parallel. Even for simple models, initially breaking down the full model into smaller individual models may speed up convergence, as observed e.g. in Figures \ref{example1_J} and \ref{fig:example2_J_10}, where parallelization alone does not explain the reduction in computation time. For more complex models, such as that of Section~\ref{sec:cheese}, the advantage is even more substantial.

The substitute hierarchical model used in combining the posterior samples into a joint inference, is typically very fast to compute. To even further improve computational efficiency, we have made extensive use of conjugacy by suitable choices of substitute likelihood functions. In Section~\ref{sec:cheese}, we showed that such conjugate forms may be utilized in the substitute model even if the original hierarchical model itself is not fully conjugate. Further theoretical analysis is, however, required to investigate the conditions and scope of this strategy. While conjugacy is not a requirement in the general methodology, we have still assumed that the observed posterior samples can be adequately characterized by \emph{some} parametric family. This is obviously an assumption which cannot always be satisfied. Therefore, extending the current approach for arbitrary posterior distributions through nonparametrics, while being able to maintain the desired computational efficiency, will be an important direction for future research.   

Although the main motivation of this paper has been to develop a parallelized strategy for inference in large and complex hierarchical models, the approach should be equally applicable, with minor or no modifications, for conducting a meta-analysis on individually conducted Bayesian analyses, where study-specific posterior samples are observed instead of the original data. Indeed, the close connection of our approach to meta-analysis was briefly discussed in Section~\ref{sec:interpretation}, and the potential for such analyses was also alluded to in the variance components model example in Section~\ref{num_illus}. In contrast to traditional meta-analysis, an approach combining individual posterior distributions need not rely on assumptions of asymptotic normality, and additionally, it allows study-specific prior knowledge to be incorporated and propagated into the analysis through the observed posterior distributions.

\begin{acknowledgements}
We acknowledge the computational resources provided by the Aalto Science-IT project.
\end{acknowledgements}

\bibliographystyle{spbasic}      
\bibliography{metanalysis_references}   


\end{document}